\documentclass[useAMS,usenatbib,aas_macros]{mn2e}

\usepackage{color}
\usepackage{epsfig}
\usepackage{lscape,graphicx}
\usepackage{amssymb}
\usepackage{natbib}
\usepackage{times}
\usepackage{aas_macros}

\newcommand{\ifm}[1]{\relax\ifmmode#1\else$\mathsurround=0pt#1$\fi}
\newcommand{\kms}{\ifmmode\,{\rm km}\,{\rm s}^{-1}\else km$\,$s$^{-1}$\fi}

\newcommand{\hmsun}{\,\ifm{h^{-1}}{M_{\odot}}}
\def\omm{\Omega_{\rm m}}
\def\oml{\Omega_{\Lambda}}

\newcommand{\ltsima}{$\; \buildrel < \over \sim \;$}
\newcommand{\lsim}{\lower.5ex\hbox{\ltsima}}
\newcommand{\gtsima}{$\; \buildrel > \over \sim \;$}
\newcommand{\gsim}{\lower.5ex\hbox{\gtsima}}

\newcommand{\dd}{{\rm d}}

\newcommand{\equ}[1]{eq.~(\ref{eq:#1})}
\newcommand{\equs}[1]{eqs.~(\ref{eq:#1})}
\newcommand{\Equ}[1]{Eq.~(\ref{eq:#1})}

\newcommand{\se}[1]{\S\ref{sec:#1}}
\newcommand{\fig}[1]{Fig.~\ref{fig:#1}}

\newcommand{\Fig}[1]{Figure~\ref{fig:#1}}

\newcommand{\dS}{\Delta S}
\newcommand{\dW}{\Delta \omega}

\newcommand{\be}{\begin{equation}}
\newcommand{\ee}{\end{equation}}
\newcommand{\bea}{\begin{eqnarray}}
\newcommand{\eea}{\end{eqnarray}}

\def\ltsima{$\; \buildrel < \over \sim \;$}
\def\lsim{\lower.7ex\hbox{\ltsima}}

\def\lan{\langle}
\def\ran{\rangle}

\def\mmin{M_{\rm min}}
\def\mmax{M_{\rm max}}
\def\macc{M_{\rm acc}}
\def\ptot{P_{\rm tot}}

\begin{document}

\title[Merger Rates of Dark-Matter Haloes]
      {Merger Rates of Dark-Matter Haloes}

\author[E.~Neistein \& A.~Dekel]
       {Eyal Neistein\thanks{E-mails: eyal\underline{$\;\;$}n@phys.huji.ac.il;
        dekel@phys.huji.ac.il}
        and Avishai Dekel$^{\star}$
        \\
      Racah Institute of Physics, The Hebrew University,
          Jerusalem, Israel }


\large

\pagerange{\pageref{firstpage}--\pageref{lastpage}} \pubyear{2008}
\maketitle

\label{firstpage}


\begin{abstract}

We derive analytic merger rates for dark-matter haloes within the framework of
the Extended Press-Schechter (EPS) formalism. These rates become self-consistent
within EPS once we realize that the typical merger in the limit of a small
time-step involves more than two progenitors, contrary to the assumption of
binary mergers adopted in earlier studies.  We present a general method
for computing merger rates that span the range of solutions permitted by the
EPS conditional mass function, and focus on a specific solution that
attempts to match the merger rates in $N$-body simulations.
The corrected EPS merger rates are more accurate than the earlier
estimates of Lacey \& Cole, by $\sim20\%$ for major mergers and by up to a
factor of $\sim3$ for minor mergers of mass ratio $1:10^4$.
Based on the revised merger rates, we provide a new algorithm
for constructing Monte-Carlo EPS merger trees, that could be useful in
Semi-Analytic Modeling. We provide analytic expressions and plot numerical
results for several quantities that are very useful in studies of galaxy
formation.  This includes
(a) the rate of mergers of a given mass ratio per given final halo,
(b) the fraction of mass added by mergers to a halo, and
(c) the rate of mergers per given main progenitor.
The creation and destruction rates of haloes serve for a
self-consistency check.
Our method for computing merger rates can be applied to conditional mass
functions beyond EPS, such as those obtained by the ellipsoidal collapse
model or extracted from $N$-body simulations.

\end{abstract}




\section{Introduction}
\label{sec:intro}

The hierarchical clustering of dark matter is the key process in
establishing the observed structure in the universe. Galaxies form
inside the potential wells induced by the dark-matter distribution.
The building blocks of this hierarchy are virialized collapsed gravitating
systems in pressure equilibrium --- the dark-matter haloes ---
characterized by their growth history, structure, and clustering.
Although dark-matter dynamics is governed solely by the gravitational force,
we are still far from a good quantitative understanding of its various
features.

The Press-Schechter (PS) formalism \citep{Press74} has been very useful
in modeling the abundance of dark-matter haloes as a function of mass
and time.  It has been further developed by \citet{Bond91} and
\citet[][hereafter LC93]{Lacey93} to the Extended Press-Schechter (EPS)
formalism, which provides at any time the mass function of progenitors
of a halo of a given current mass. EPS has been a basic tool for
understanding the growth history of haloes, and it has been shown to
grasp many of the key features of the buildup of haloes in cosmological
$N$-body simulations \citep[e.g.][]{Lacey94,Cole08,Neistein08}.
While EPS has been used extensively for the last two decades,
it still involves central open issues. One is the construction of
self-consistent Monte-Carlo merger trees for Semi-Analytic Models
of galaxy formation. The other is how to compute halo merger rates
that will be consistent with the EPS mass function.

While drawing the basic lines of the EPS theory, LC93 worked out a formula
for the merger rates of haloes. This formula has been popular in many
applications, although it involves a problem. LC93 themselves noticed that
their merger-rate formula has a problematic intrinsic asymmetry between
progenitors of mass $M$ and $M_0-M$ (where $M_0$ is the descendant halo mass).
\citet{Sheth97} realized that the LC93 assumption of binary mergers
is not accurate when the power spectrum differs from a white noise
(their discussion near eq. 27). \citet{Benson05} interpreted this as an
intrinsic inconsistency within the EPS formalism.
We show below that the typical mergers have multiple progenitors, more than
two, even in the limit of a small time step. Adopting this correct limit, we
obtain accurate EPS merger rates, which improve the LC93 estimates
and are fully consistent with the EPS conditional mass function.
The error in the LC93 formula makes a significant difference
for the number of merger events and for the fraction of halo mass
added by mergers.

Random realizations of merger trees that follow the EPS conditional
mass function are widely used as the back-bone of semi-analytic modeling
of galaxy formation. Several different methods for constructing such
trees have been proposed
\citep{Cole91, Kauffmann93, Sheth99, Somerville99a, Cole00, Hiotelis06}.
In most cases, these algorithms fail to recover the EPS mass function.
It seems that the algorithm of \citet{Kauffmann93} is the only one that is
fully consistent with EPS. \citet{Sheth99} described an alternative that is
also accurate, but it has not been developed into a detailed solution.
One can indeed show that the EPS formalism permits many
different types of merger trees that recover the EPS progenitor mass
function.  We provide below a new algorithm for constructing EPS merger trees
based on our formula for merger rates. This algorithm does reproduce the EPS
progenitor mass function, and it is chosen among the different solutions
to be a good match to the merger trees extracted from cosmological $N$-body
simulations.

Empirical algorithms for generating merger trees that resemble
the trees in cosmological $N$-body simulations have been proposed
by \citet{Parkinson08,Neistein08}. These merger trees are for most
parts better approximations to the $N$-body results than any EPS-based
tree. However, a correct EPS model has several useful benefits. For
example, it allows very high mass resolution at low cost, it can be
easily applied within any desired cosmological model, and it is
self-consistent with the Press-Schechter halo abundance.
On the other hand, the empirical algorithms mentioned above should
always be verified and re-calibrated when used in a different
cosmology or when applied at a different resolution.
Analytic models in the spirit of EPS can serve us in understanding several
open issues concerning the way haloes are identified in $N$-body
simulations. For example, it has been noticed
\citep[][hereafter ND08]{Neistein08}
that some of the non-Markov features in $N$-body merger
trees may arise from the way haloes are defined. Indeed, the halo
definition has become an open issue with the finding that the range
of virial equilibrium in small haloes can extend well beyond the
traditional ``virial radius" that is based on  spherical collapse
\citep{Cuesta07,Ludlow08}. As part of this work we provide
a general method for generating merger trees that follow any given
conditional mass function. This mass function could be either based
on spherical collapse (i.e., EPS), or arise from ellipsoidal collapse
\citep{Sheth02}, or extracted from $N$-body simulations.

This paper is organized as follows. In \se{general} we present
nomenclature, describe the limit of small time-steps, and prove
the theorem concerning multiple progenitors. In \se{specific_solut}
we address different solutions for the EPS halo merger rates, and
choose the solution that fits well the $N$-body results. In \se{implic}
we work out useful results for merger rates from our EPS formalism,
and present them in practical formulae and in figures.
In \se{halo_creation} we address the creation and destruction rates of
haloes.
In \se{algo} we describe a Monte-Carlo algorithm for constructing
EPS merger trees based on our adopted solution. In \se{discuss} we
summarize our results and discuss them.


\section{General Analysis}
\label{sec:general}

\subsection{Definitions: PS and EPS}
\label{sec:definition}

In the EPS formalism, the natural dimensionless time variable is
$\omega(z) = \delta_c(z)/D(z)$, where $D(z)$ is the cosmological
linear growth rate of density fluctuations as a function of redshift
$z$ and $\delta_c\simeq1.69$. The natural mass variable is
$S(M)=\sigma^2(M)$, the variance of the initial density fluctuation
field, linearly extrapolated to $z=0$, and smoothed using a window
function that corresponds to a mass $M$. The reader is referred to
ND08 for our specific way for computing these quantities. The cosmological
model used here is defined by
$(\oml, \, \omm,\,h, \,\sigma_8 )= (0.75, \, 0.25,\, 0.73, \, 0.9)$,
with the power spectrum specified in ND08. This model was adopted to enable
comparison with results extracted from the Millennium cosmological simulation
\citep{Springel05}.

According to the EPS formalism \citep[][LC93]{Bond91}, the average number
of progenitors in the mass interval $[M, M+\dd M]$, which will merge
into a descendant halo $M_0$ after a time-step $\dW$, is given by
\begin{eqnarray}
\label{eq:dNdM}
\lefteqn{ {{\dd}N \over {\dd}M}(M \vert M_0,\dW) \,\dd M
\; = } \\ \nonumber & & {M_0 \over M} \; \frac{1}{\sqrt{2 \pi}} \;
{\dW \over (\dS)^{3/2}}
\; {\rm exp}\left[-{(\dW)^2 \over 2 \dS}\right] \left\vert
\frac{\dd S}{\dd M} \right\vert \,\dd M \;   \,,
\end{eqnarray}
where $\dS=S(M)-S(M_0)$. We term the most massive progenitor in this
time-step by $M_1$, the second most massive by $M_2$, and so on.
The probability that $M$ is the mass of the $i$-th progenitor is termed
$P_i=P_i(M |M_0,\dW)$. Consequently, the sum of all the $P_i$'s equals
$\dd N/\dd M$:
\begin{eqnarray}
\ptot (M|M_0,\dW) &\equiv& \frac{\dd N}{\dd M}(M|M_0,\dW) \\
\nonumber &=& \sum_i P_i(M|M_0,\dW) \;.
\end{eqnarray}
For brevity, we may sometimes omit the explicit dependence of $\ptot$ and
$P_i$ on $M_0$ and $\dW$.

It is often useful to define a minimum halo mass, $\mmin$. Haloes with
smaller masses are considered to be part of a smooth accretion component,
encompassing a total mass $\macc$.

We also need the total number density of haloes per unit mass per
comoving volume, which is given by the Press-Schechter mass function:
\begin{equation}
\label{eq:PS}
\phi(M,z) = \frac{1}{\sqrt{2\pi}} \frac{\rho_{_0}}{M}
\frac{\omega }{S^{3/2}}
\exp\left[-\frac{\omega^2 }{2S}\right] \left| \frac{\dd S}{\dd M}
\right| \;,
\end{equation}
where $\rho_{_0}$ is the present mean mass density of the universe.


\subsection{Number of Progenitors in a Small Time-Step}
\label{sec:time-step}

Throughout this work, we appeal to the limit of a small time-step,
$\dW\!\rightarrow\!0$, relevant for the derivative with respect to ``time",
$\dd/\dd\omega$. For given $M_0$ and $\mmin$, the limit of a small
time-step is defined here as $\dW\ll S(M_0-\mmin) - S(M_0)$. In this
limit, and when $M\leq M_0-\mmin$, the probability $\ptot$ can be written as
\begin{eqnarray}
\label{eq:ptot0}
\ptot(M|M_0,\dW\rightarrow0) =
\frac{1}{\sqrt{2\pi}} \frac{M_0}{M}
\frac{\dW}{(\dS)^{3/2}} \left| \frac{\dd S}{\dd M} \right| \;,
\end{eqnarray}
after the exponent in eq.~\ref{eq:dNdM} has been set to unity.
Consequently, the ``time" derivative of $\ptot$ is simply
\begin{eqnarray}
\label{eq:dptot_dw}
\frac{\dd \ptot(M|M_0)}{\dd\omega} = \frac{1}{\sqrt{2\pi}}
\frac{M_0}{M} \frac{1}{(\dS)^{3/2}} \left| \frac{\dd S}{\dd M} \right| \;.
\end{eqnarray}
We occasionally write $\dd/\dd\omega$ when it should formally be
$\dd/\dd\dW$, as both derivatives are the same\footnote{We assume that
$\dW=\omega-\omega_0$ and the derivative $\dd/\dd\omega$ is computed at
a fixed $\omega_0$.}. The above equations are valid only for $M\leq M_0-\mmin$;
otherwise $\dS$ may also become infinitely small, such that $(\dW)^2/\dS$
does not vanish, and the exponent in \equ{dNdM} does not converge to unity.

\begin{figure}
\centerline{ \hbox{ \epsfig{file=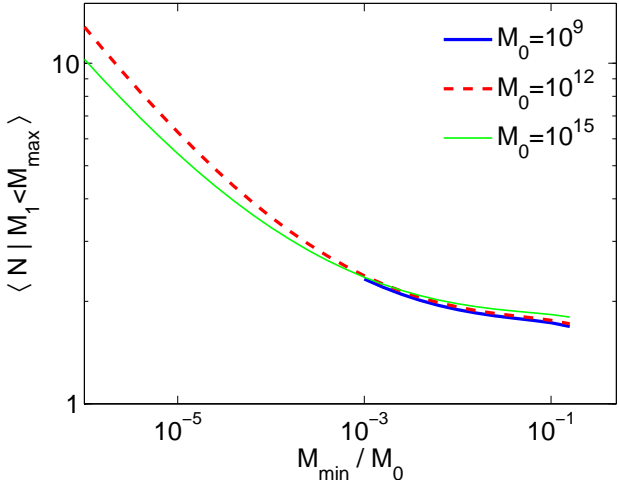,width=9cm} }}
\caption{The average number of progenitors given that the main progenitor
mass is less than $\mmax\!=\!M_0-\mmin$, as a function of $\mmin/M_0$.
The three different curves are for different values of $M_0$
as indicated (with units of $\hmsun$).
Each of the curves is plotted only for $\mmin>10^6\;\hmsun$.
The computation is done in the limit of a small time-step,
$\dW\!\rightarrow\!0$. Evidently, if the minimum mass is less than
$\sim 10^{-3}M_0$, the number of progenitors is larger than two.
This implies that the concept of binary mergers is highly inaccurate for low
values of $\mmin/M_0$.
This conclusion is valid independently of the value of $M_0$.
}
  \label{fig:n_progs}
\end{figure}

We now prove the theorem of multiple progenitors, claiming that
according to EPS, the typical merger involves multiple progenitors
rather than a binary merger even in the limit of a small time-step.
{\bf Theorem: Given the EPS progenitor mass function of \equ{dNdM},
with the CDM power-spectrum, in the range
$\bf \mmin\!\ll\! M_0$ and in the limit $\bf \dW \!\rightarrow\! 0$,
the average number of progenitors per merger event is greater than two.}

We first notice that the constraint of mass conservation, that the total
mass in progenitors cannot exceed $M_0$, implies that events with
$M_1\!>\!\mmax$, where $\mmax\equiv M_0\!-\!\mmin$, cannot have any other
progenitor with $M_i\!>\!\mmin$. Therefore, merger events between two or
more progenitors above $\mmin$ are limited to the cases where $M_1\!<\!\mmax$.

Let $N$ be the number of progenitors with mass in the range $[\mmin,\mmax]$.
We first show that $\lan N | M_1\!<\!\mmax \ran > 2$. If $P(M_1\!<\!M)$ is
the probability that $M_1\!<\!M$, then $\lan N \ran = P(M_1\!<\!\mmax) \times
\lan N | M_1\!<\!\mmax \ran$, because the contribution of the other events
is zero progenitors\footnote{Strictly speaking, we should write
$\lan N \ran = P(\mmin\!<\!M_1\!<\!\mmax) \times
\lan N | \mmin\!<\!M_1\!<\!\mmax \ran$. Using the fact that
$P(\mmin\!<\!M_1\!<\!\mmax)\leq P(M_1\!<\!\mmax)$ it is evident that
all the results proved here using $P(M_1\!<\!\mmax)$ are lower limits on
the accurate $\lan N | \mmin\!<\!M_1\!<\!\mmax \ran$.}.  We thus obtain
\begin{eqnarray}
\label{eq:nprog}
\lan N | M_1\!<\!\mmax \ran
&=& \frac{\lan N \ran}{P(M_1\!<\!\mmax)}\\ \nonumber
&=& \frac{ \int_{\mmin}^{\mmax} \ptot(M) \dd M }
{1 - \int_{\mmax}^{M_0} P_1 (M) \dd M } \;.
\end{eqnarray}
When we calculate the integral in the denominator, we note that $P_1$
can be replaced by $\ptot$ near $M_0$ (see the discussion preceding
eq.~\ref{eq:p1_default} below). As $\dW \rightarrow 0$,
\equ{ptot0} implies that $\lan N \ran$ vanishes in proportion to $\dW$,
but $P(M_1\!<\!\mmax)$ also vanishes\footnote{When computing the denominator
one should use eq.~\ref{eq:dNdM}, and not the approximation of eq.~\ref{eq:ptot0}.
In the case where $\mmin\ll M_0$ we can approximate $M_0/M\sim1$ and the
denominator is just ${\rm Erf}[\dW/\sqrt{2\dS_m}] \sim  \sqrt{2/\pi}\dW/\sqrt{\dS_m}$, where $\dS_m=S(\mmax)-S(M_0)$ and $\dW\rightarrow0$. When $\mmin$ is not small enough, the integral in the denominator can be
computed only numerically.},
making the ratio converge to a finite value.

\Fig{n_progs} shows the average number of progenitors given that the
main-progenitor mass is smaller than $\mmax$, $\lan N | M_1\!<\!\mmax\ran$.
This is in the limit of a small time-step and for different values of
$M_0$. We see that this average is greater than two for any
$\mmin<10^{-3}M_0$. It increases with decreasing $\mmin$ to a value of
$\sim\!10$ for $\mmin=10^{-6}M_0$. This proves that
$\lan N | M_1\!<\!\mmax \ran > 2$.

Since each of the events with $M_1\!<\!\mmax$ that are not mergers
contributes to the conditional average of $N$ a value $\leq 1$, the
merger events, which are a subset of the $ M_1\!<\!\mmax$ events,
must have on average even more progenitors than computed in \equ{nprog}
and shown in \fig{n_progs}. We conclude that the assumption of binary
mergers is invalid in EPS, even for $\dW \!\rightarrow\! 0$, once
$\mmin\! <\! 10^{-3} M_0$. This proves the theorem.

If $M_0$ is not that much larger than $\mmin$, the range
$\mmin \! \geq \! 10^{-2} M_0$ in \fig{n_progs}, we obtain
$\lan N | M_1\!<\!\mmax \ran \lsim 2$. This implies that the average
mass of the two progenitors does not sum up to $M_0$, namely the
accretion component $\macc$ contains a non-negligible fraction of the
mass.

Since the theorem of multiple progenitors has interesting implications
on the formation of structure, it would be worthwhile to consider analytically
the average number of progenitors in the idealized case where the power
spectrum is a pure power law, $S\propto M^{-\alpha}$. Solving for $\lan N |
M_1\!<\!\mmax\ran$, we find that it is bigger than 2 for any $0<\alpha<1$
(once $\mmin$ is small enough). For $\alpha=1$, the case of Poisson white
noise, one can show that $\lan N | M_1\!<\!\mmax\ran \rightarrow 2$ when
$\mmin/M_0 \rightarrow 0$,
in agreement with the coagulation approach discussed by \citet{Epstein83}
and \citet{Sheth97}.
For $\alpha>1$, the average number of progenitors never exceeds 2.
We learn that the average number of progenitors per merger event depends on
the shape of the power spectrum. In particular, for power spectra that are
relevant on galactic scales, $\alpha < 1$, the average number of progenitors
per merger are more than two.

We note that the existence of multiple mergers in the limit of small time-steps
is already mentioned in \citet{Sheth97}. \citet{Sheth99} added that
for a general power spectrum, one can group the progenitors into sub-groups
that merge like the progenitors of the Poisson-power-spectrum case.


\subsection{Merger rates}
\label{sec:merger-rates}

One way to define a merger rate is as the probability for the $i$-th
most massive progenitor to merge into the main progenitor within a
time-step $\dW$. This is the joint probability for the two progenitor
masses $M_1$ and $M_i$, which we denote $P_{1,i}(M_1,M_i|M_0,\dW)$.
Note that $P_i$ and $P_j$ can both have non-vanishing values at the same
mass, so the probability for \emph{any} progenitor with mass $M_s$ to merge
with $M_1$ is the sum
\begin{equation}
P_{1,s}(M_1,M_s|M_0,\dW) \equiv \sum_i P_{1,i}(M_1,M_s|M_0,\dW) \;.
\end{equation}

We learned in \se{time-step} that there are typically several progenitors
in each merger event even in the limit of a small time-step. To complicate
matters even further, we note that $P_{1,i+1}$ is not necessarily smaller
than $P_{1,i}$. Still, for the purpose of estimating merger rates, we wish
to approximate this multi-progenitor merging process as an instantaneous
sequence of binary mergers. There is clearly no unique way to do that. We
adopt here the assumption that each of the secondary progenitors ($M_i, i>1$)
merges with a halo of mass $M_1$, and ignore mergers among the secondary haloes
themselves. The validity of this assumption can be tested in $N$-body
simulations\footnote{In \se{discuss}, we discuss the possible relation between 
a multiple merger event in EPS and a correlated sequence of binary
mergers in an $N$-body simulation.
If the time between mergers in the $N$-body sequence is shorter
than the time it takes the remnant halo to settle into its new potential, 
than our assumption might be reasonable.}.
This assumption makes sense when $M_1$ is much more massive than the other
progenitors. However, in a case where $M_1\sim M_2 \gg M_3$, one might consider
$M_3$ merging with a halo of mass $M_1+M_2$ instead. We assume that this
uncertainty in interpreting the multiple merger events does not translate to
a significant error in our estimated average merger rate, but the actual
estimate of this uncertainty is beyond the scope of the present paper.

For any progenitor $M_i$, we define $P_{i|1}(M_i|M_1,M_0,\dW)$ to be the
conditional probability to have $M_i$ \emph{given} that the main-progenitor
mass is $M_1$. Then
\begin{eqnarray}
\lefteqn{ P_{1,i}(M_1,M_i|M_0,\dW) = } \\
\nonumber & & P_{i|1}(M_i|M_1,M_0,\dW) \; \cdot P_1(M_1|M_0,\dW) \;.
\end{eqnarray}
Our approach for finding a solution for $P_{1,i}$ starts with
a solution for $P_1$, followed by a solution for $P_{i|1}$.
This is because $P_1$ is determined robustly by EPS, with only a small,
controllable uncertainty over a limited mass range.

\begin{figure}
\centerline{ \hbox{ \epsfig{file=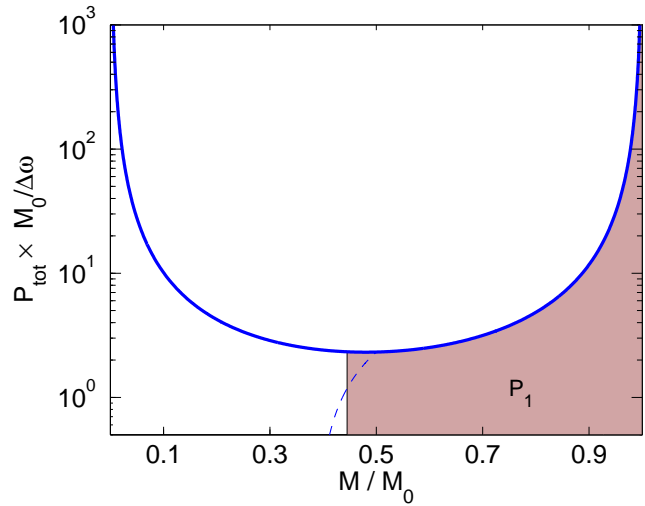,width=9cm} }}
\caption{Two possible solutions for the probability distribution
of the main progenitor, $P_1$.
The thick blue curve corresponds to $\ptot$, properly normalized
as indicated with $M_0=10^{12}\;\hmsun$ and $\dW=10^{-6}$.
The solutions for $P_1$ differ only in the small tail at $M \lsim M_0/2$.
The shaded area marks the range over which the integral of
$\ptot$ equals unity; it ends at $x_1 =M/M_0 \sim 0.44$.
This is also our default definition of $P_1$, termed \emph{sharp tail}.
The dashed curve marks another possible tail, also corresponding
to an integral of unity, which is linear in $M$, and thus termed
\emph{linear tail}. Note that we plot $\ptot\times M_0/\dW$ as this curve
is the same for all small $\dW$, in accord with eq.~\ref{eq:ptot0}.
}
  \label{fig:P_tot}
\end{figure}

The shape of $P_1$, the small freedom in it within EPS,
and its effect on the average mass history of the main progenitor
has been studied in \citet{Neistein06} and can be summarized as follows:
In the range $M_1\geq M_0/2$, $P_1$ is identical to the known $\ptot$,
because any progenitor in this mass range is by definition the main
progenitor. For $M_1$ slightly below $M_0/2$, there is a ``tail" of
non-vanishing probability, which could go to zero in many different ways.
This is illustrated in \fig{P_tot}, which shows two of the many possible
solutions for this tail. Our default option is with a `sharp tail',
\begin{equation}
\label{eq:p1_default}
P_{1}(M_1|M_0,\dW) = \left\{ \begin{array}{ll}
\ptot & \textrm{if $M_1>x_1M_0$}\\ \;\;\;\
& \;\;\;\;\;\;\;\;\;\;\;\;\;\;\;\;\;\;\;\;\;\;\;\;\;\;\;\;\;\;\;\;\;\;\; . \\
0 & \textrm{if $M_1\leq x_1M_0$}
\end{array} \right.
\end{equation}
The value of $x_1$ is set by the requirement that the integral over $P_1$
equals unity. It is $\simeq 0.44$ for the cosmology and for the halo masses
used here\footnote{{Using $u=\log_{10}(M_0)-12$ where $M_0$ is in units of
$\hmsun$ we can approximate} $x_1 =
7.118\times10^{-5} u^3 + 6.225\times10^{-4} u^2 + 0.0035u + 0.444$ with an
accuracy that is better than $0.05\%$.}. We note that the average mass
history of the main-progenitor using this $P_1$ can be computed by the
analytical formula of \citet{Neistein06}. Fig.~\ref{fig:P_tot} also shows
an alternative solution where the $P_1$ tail is linear in $M$. The freedom
in the tail of $P_1$ corresponds to an uncertainty of less than 8\% in the
average relative growth rate of the main progenitor, $(\dd M_1/\dd\omega)/M_1$.

Assuming a specific solution for $P_1$, the constraints for having a correct
$P_{i|1}$ are as follows:
\begin{eqnarray}
\lefteqn{ P_i(M_i|M_0) =  \int P_{i|1}(M_i|M_1,M_0) \; P_1(M_1|M_0)
\; \dd M_1}
\label{eq:pi1_constraint1} %
\\
\lefteqn{ \ptot(M) = \sum_i P_i(M) }
\label{eq:pi1_constraint2} %
\\
\lefteqn{ P(M_1,M_2,\ldots) = 0 \;\;\;\; {\rm if } \;\;\;\; \sum_i M_i>M_0 }
\label{eq:pi1_constraint3} %
\end{eqnarray}
The last condition is assuring mass conservation, where the total
mass of all progenitors cannot be larger than $M_0$.

For certain purposes, it will be helpful to define $P_{s|1}$ as the
sum over all $P_{i|1}$. The constraint for $P_{s|1}$ is simply
\begin{eqnarray}
\label{eq:ps1_constraint}
\ptot(M) - P_1(M) = \int P_{s|1}(M|M_1) P_1(M_1) \dd M_1 \;.
\end{eqnarray}
Here mass conservation cannot be formulated as an explicit condition
on $P_{s|1}$ because it does not contain information concerning
the mutual distribution of multiple progenitors.

While $P_1$ is robustly determined in EPS, there is a great deal of
freedom in $P_{1,i}$. This is because $P_{1,i}$ is a two-dimensional
function with only one-dimensional constraints \citep[e.g.][]{Benson05}.
We emphasize that this is true also for small time-steps. Hence there
are many solutions for the desired EPS merger rates. In the next section
we show several valid solutions of this sort.


\section{Specific Solutions}
\label{sec:specific_solut}

Here we bring a general formalism for obtaining solutions
$P_{1,i}$ and demonstrate the level of freedom allowed while obeying
the EPS conditional mass functions.
Given the robust expression in \equ{ptot0} for $\ptot$ in the limit of
a small time-step, the solutions presented below are valid for any value
of $\dW$ once it is small enough.


\subsection{Determining a unique set of $M_i$'s for a given $M_1$}
\label{sec:fixed}

Our general solution is motivated by the merger-rate concept introduced
by LC93. Assume that for any $M_1$ we can choose a \emph{unique} set of
smaller progenitors $\{M_i\}$, so that each $P_{i|1}$ is a delta function:
\begin{eqnarray}
\label{eq:define_f}
P_{i|1}(M_i|M_1,M_0,\dW) = \delta \left[ M_i - f_i(M_1|M_0,\dW) \right] \;.
\end{eqnarray}
Here $f_i(M_1|M_0,\dW)$ associates a value of $M_i$ to any $M_1$. We often
write $f_i(M_1)$ where $M_0$ and $\dW$ are obvious from the context.
Substituting $P_{i|1}$ from \equ{define_f} in the constraint of
\equ{pi1_constraint1}, and integrating over $M_1$, we obtain a differential
equation for $f_i(M_1)$:
\begin{eqnarray}
\label{eq:fi_ode}
\frac{\dd f_i(M_1)}{\dd M_1} =  - \frac{P_1(M_1)}{P_i \left[ f_i(M_1) \right] } \;,
\end{eqnarray}
where $f_i$ is assumed to be a monotonically decreasing function of $M_1$.
Thus, the solution for $f_i(M_1)$ is determined by $P_1$, $P_i$, and a
certain initial condition $M_{i,0}=f_i(M_{1,0})$. This differential equation
is to be integrated numerically to obtain a solution for $f_i(M_1)$. Note,
in contrast, that LC93 adopted the inaccurate assumption $f_2^{LC}(M_1)=M_0-M_1$,
failing to allow for the additional progenitors beyond $M_2$.

\begin{figure}
\centerline{ \hbox{ \epsfig{file=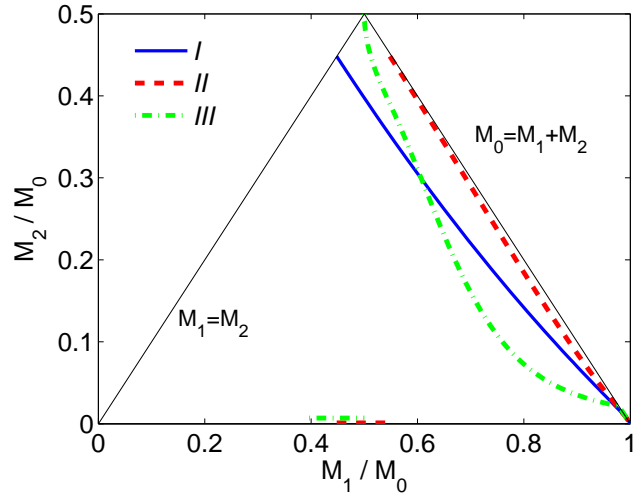,width=9cm} }}
\caption{Three solutions for $P_{1,2}$,
with a unique $M_2$ for each $M_1$.
The solutions are derived here for $M_0=10^{13}\;\hmsun$, $\dW=10^{-6}$;
they are practically the same for any smaller $\dW$.
The solid (blue) and dashed (red) curves are computed for the same
$P_1$ (the default sharp tail), and they differ only in the initial
conditions (termed solutions $I$ and $II$).
The dotted-dashed (green) curve is obtained using the linear tail for $P_1$
(solution $III$). Note that the dashed and dotted-dashed lines have
disconnected segments near $(M_1,M_2)\sim(M_0/2,0)$.
The solid line is our default solution ($I$).
A summary of these solutions can be found in table \ref{tab:solutions}.}
  \label{fig:P_12}
\end{figure}

We start, for example, with $P_{2|1}$, using our default sharp-tail
solution for $P_1$ as in \equ{p1_default}. Given this $P_1$, we
try to set $P_2=\ptot-P_1$, which simply equals $\ptot$ in the range
$M<x_1M_0$. A solution for $f_2(M_1)$ can now be obtained for a given
initial condition.  Our first choice, which we term ``Solution $I$", is
\begin{equation}
\label{eq:default_IC}
(M_{1,0},M_{2,0})=(x_1M_0,x_1M_0) \;.
\end{equation}
This ensures that $M_2$ approaches $M_1$ as the latter obtains its minimum
value $x_1M_0$. Solution $I$ is shown in \fig{P_12}. We also plot the solution
for the initial condition $(M_{1,0},M_{2,0})=(M_0-x_1M_0,x_1M_0)$, termed
solution $II$. As is evident from the figure, although both solutions have the
same $P_1$, they have quite different values of $P_{1,2}$. \Fig{P_12} also
shows Solution $III$, which is based on $P_1$ with the ``linear tail" shown
in \fig{P_tot}. A summary of these three solutions is listed in
table~\ref{tab:solutions}. It should be noted that solutions $II$ and $III$
include a small range of $M_1$ values near $M_0/2$ that is not connected to
$M_2$ through $M_2=f_2(M_1)$. For $M_1$ in this range we
cannot use any value of $M_2$ that was previously associated with
$M_1$ through $f_2$. We can either choose $M_2\equiv0$,
or treat it similarly to $P_3$, as will be explained below.

Luckily, the condition for mass conservation is almost fully obeyed by each
of the three solutions for $f_2(M_1)$ above. This is implied by the fact that
the curves for $P_{1,2}$ seem to always lie below the line $M_1+M_2=M_0$.
However, a closer look shows that this constraint is violated for
$M_1\gtrsim0.99M_0$. In this range $f_2(M_1)>M_0-M_1$ for all the solutions
presented here, and we cannot adopt the $M_2$ that solves the differential
equation. Instead, we enforce $M_2=M_0-M_1$, which makes the distribution
$P_2$ differ slightly from $\ptot$. This result is expected based on the
multiple-progenitor theorem of \se{time-step}, requiring more than two
progenitors for reproducing $\ptot$.

\begin{figure}
\centerline{ \hbox{ \epsfig{file=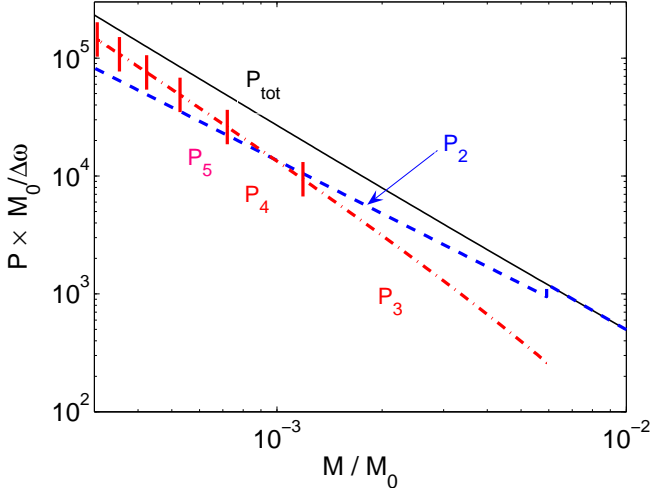,width=9cm} }}
\caption{The probability distribution for small progenitors
according to our algorithm, \equs{Pi_large} and (\ref{eq:IC_large}),
using $M_0=10^{13} \;\hmsun$ and $\dW=10^{-6}$.
The solid line is $\ptot$. The dashed (blue) curve is $P_2$.
It deviates from $\ptot$ for $M\lesssim 6\times10^{-3}\times M_0$.
The Dashed-dotted curves correspond to $P_i$ for $i>2$.
They equal $\ptot-P_2$. The vertical bars mark the limits for
each $P_i$, termed $M_{{\rm high},i}$ and $M_{{\rm low},i}$.}
  \label{fig:P_small}
\end{figure}

Fig.~\ref{fig:P_small} shows $\ptot$ and $P_2$ for small $M$ values.
The effect discussed above leads to $P_2 < \ptot$ at $M_2 < 0.01M_0$,
meaning that additional progenitors are needed in order to obtain an accurate
fit for $\ptot$.

Next we should address $P_i$ for for the $i$-th progenitor, $i>2$.
In what follows we use as an example the solution $I$ of $P_{1,2}$, and
the procedure can be easily generalized to deal with the other solutions.
For $i>2$ we define
\begin{equation}
\label{eq:Pi_large}
P_i(M) = \left\{\begin{array}{ll}
\ptot(M) - P_2(M) & M_{{\rm low},i} \!\leq\! M \!<\! M_{{\rm high},i}\\ \;\;\;\
& \;\;\;\;\;\;\;\;\;\;\;\;\;\;\;\;\;\;\;\;\;\;\;\;\;\;\;\;\;\;\;\; , \\
0 & \textrm{otherwise}
\end{array} \right.
\end{equation}
where $M_{{\rm high},i}=M_{{\rm low},i-1}$ for $i>3$ and $M_{{\rm high},3}$ is
the maximum $M$ for which $P_2<\ptot$. The value of $M_{{\rm low},i}$ is set by
the condition of mass conservation: each solution $f_i(M_1)$ is defined up to
the point where $M_0=M_1+\sum f_i(M_1)$. The initial condition is thus
\begin{equation}
\label{eq:IC_large}
(M_{1,0},M_{i,0})=(x_1M_0,M_{{\rm high},i}) \;,
\end{equation}
and the set of $P_i$ we obtain is given in fig.~\ref{fig:P_small}.

To summarize, our solution for the merger rate is
\begin{eqnarray}
\label{eq:P_1i}
\lefteqn{P_{1,i}(M_1,M_i|M_0,\dW) = } \\ \nonumber & & P_1(M_1|M_0,\dW) \;
\delta \left[ M_i - f_i(M_1|M_0,\dW) \right]  \;.
\end{eqnarray}
In the limit of a small time-step, $f_i$ does not depend on $\dW$
and $P_1$ is given by \equ{ptot0}.

\begin{table}
\caption{The characteristics of the EPS solutions for
$P_{1,2}$ discussed in \se{fixed}. The threes solutions assume the
delta-function form for $P_{2|1}$, \equ{define_f}.}
\begin{center}
\begin{tabular}{lccc}
\hline Solution & $P_1$ tail & $P_{1,2}$ Initial conditions  \\
\hline
$I$ & Sharp & $(x_1M_0,x_1M_0)$  \\
$II$ & Sharp & $(x_1M_0,1-x_1M_0)$ \\
$III$ & Linear & $(M_0/2,M_0/2)$  \\
\hline
\end{tabular}
\end{center}
\label{tab:solutions}
\end{table}
%


\subsection{Comparison with $N$-body results}

We now wish to compare the merger-rates from our EPS analysis to merger-rates
that were extracted from $N$-body simulations. We remind the reader that when
using $N$-body simulations with small time-steps, the merger-rates suffer
from inconsistencies due to non-Markov features (ND08), so any Markov model,
as the one
implied by EPS, will have deviations at small time-steps. A more fair comparison
should be done against the model derived by ND08, which should be close to the
optimum Markov fit to the simulations.

\begin{figure}
\centerline{ \hbox{ \epsfig{file=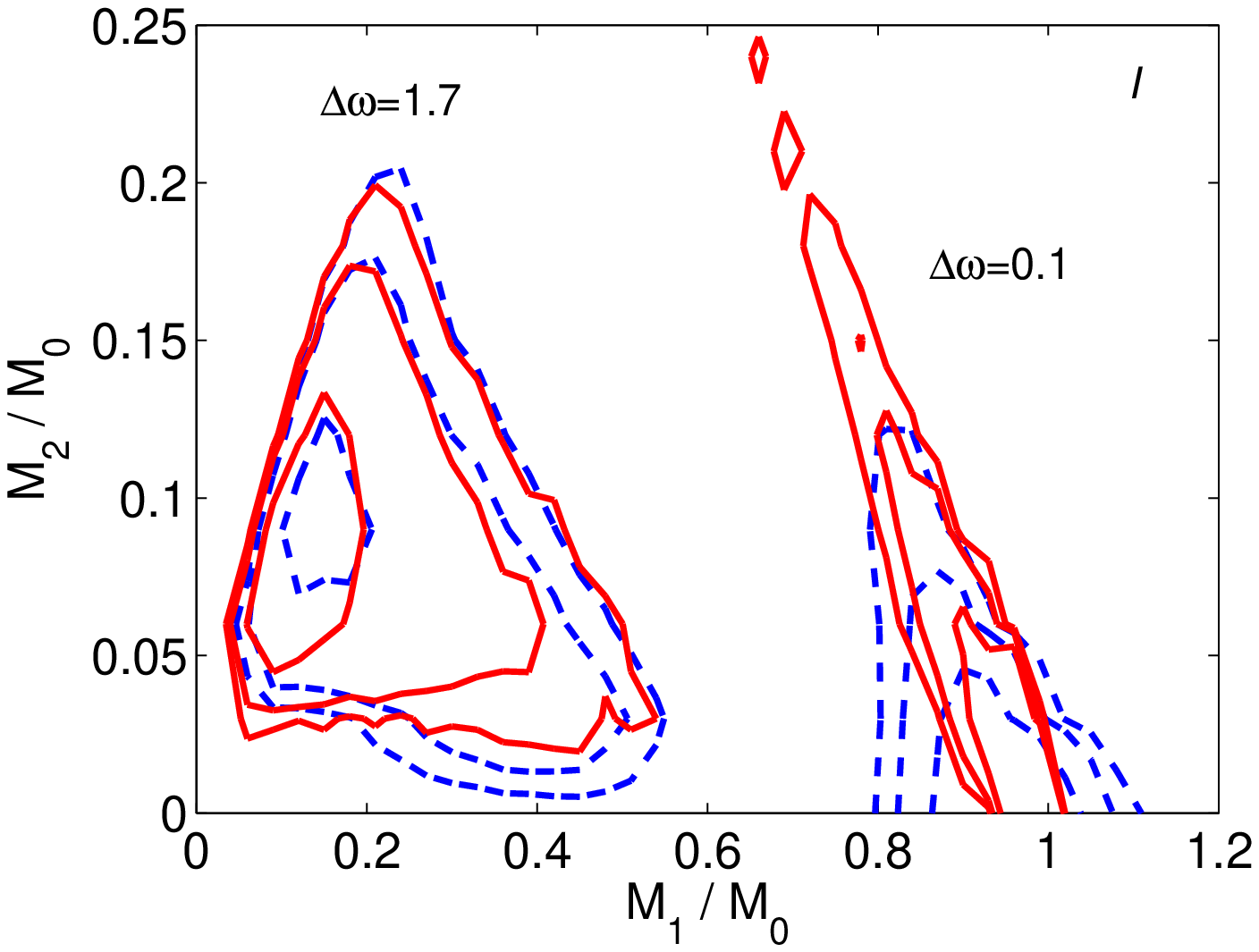,width=9cm} }}
\centerline{ \hbox{ \epsfig{file=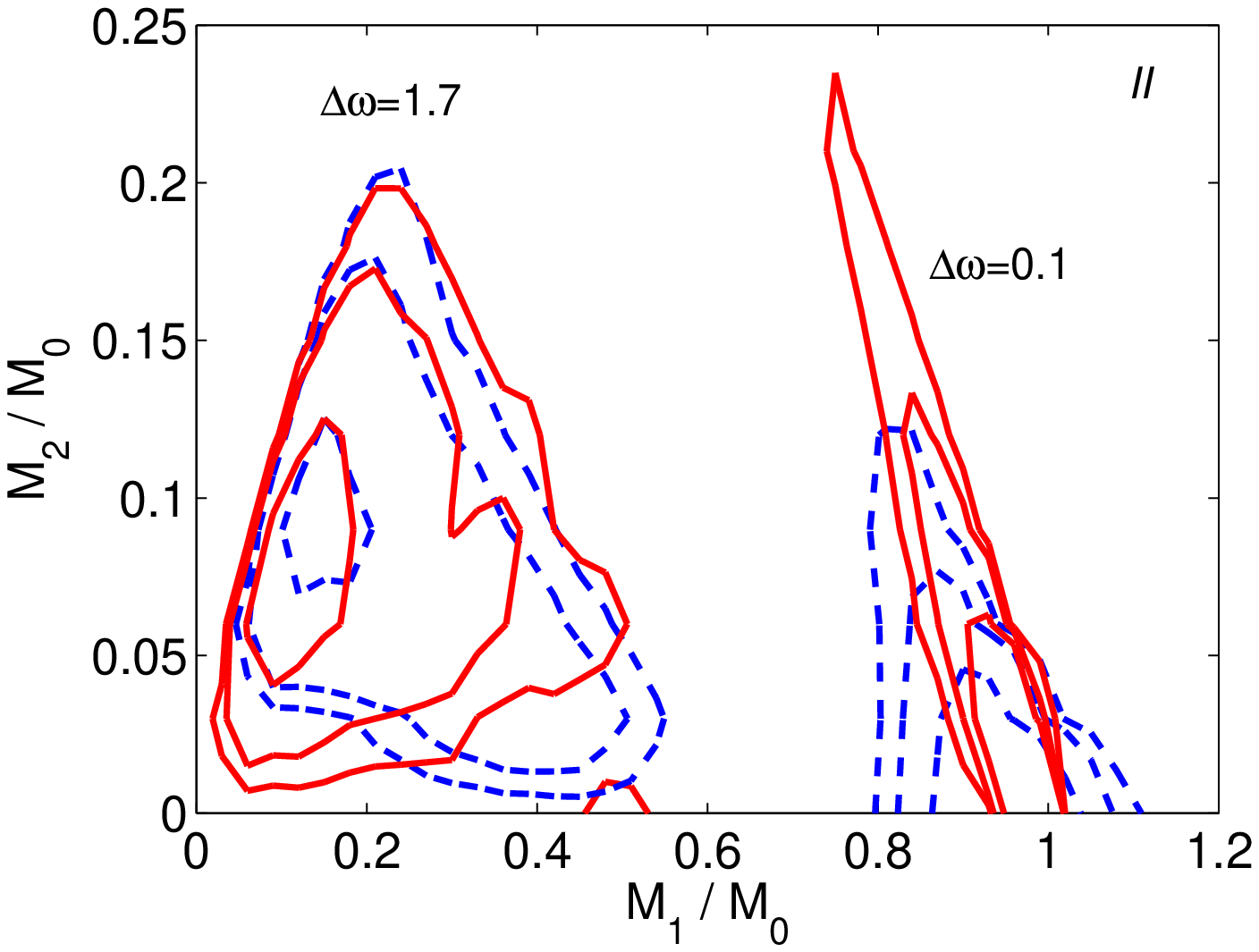,width=9cm} }}
\caption{The joint probability of the two most
massive progenitors, $P_{1,2}$, for haloes of mass $2\times10^{13}\;\hmsun$.
The plots refer to two different time-steps, $\dW=0.1$ and $1.7$.
The contour levels are at $P_{1,2}=5,10,30\; M_0^{-2}$.
\emph{Upper panel:} The solid (red) contours refer to results of the EPS
algorithm $I$, obtained from generated random realization with
intrinsic $\dW=0.001$.
\emph{Lower panel:} Same for solution $II$.
The corresponding results from the Millennium $N$-body simulation
are shown for comparison as dashed (blue) contours (previously
presented in ND08, Fig.~7).
}
  \label{fig:P12_Nbody}
\end{figure}

In \fig{P12_Nbody} we show results of our solutions $I$ and $II$ against
$N$-body simulations \citep[the Millennium run, ][]{Springel05}. In
order to test our solutions we use the merger-tree algorithm as described
in section \ref{sec:algo} below.
Fig.~7 of ND08 indicates that the EPS merger rates found here do resemble relatively well the merger rates of the Markov model that fits the simulation in ND08.
At bigger time-steps, we see that although the general contour shape is similar,
the average mass of the second progenitor is slightly smaller in EPS than in the
simulation. This is also evident in the results of \citet{Parkinson08}, who
compared a different set of $N$-body merger trees with a binary-merger model for
EPS trees a la LC93. On the other hand, merger trees constructed using the
algorithm of \citet{Somerville99a} have a significantly lower mass for $M_2$,
as pointed out in ND08. We find that the algorithm proposed by \citet{Kauffmann93} produces EPS trees that match the $N$-body trees at a level comparable to our EPS
solution $I$, though it may not be as useful as our algorithm in generating a
statistical sample of merger trees and in allowing analytic estimates.

When comparing Fig.~7 of ND08 to \fig{P12_Nbody} here, it seems that
the most important difference lies in the shape of the main-progenitor
distribution. We recall that according to ND08, for trees extracted from
$N$-body simulations, this distribution is log-normal in $S$.
On the other hand, in EPS this distribution is given by \equ{p1_default},
which has quite a different shape (see also \fig{P_tot}).

\Fig{P12_Nbody} also compares solutions $I$ and $II$,
showing that solution $I$ is somewhat closer
to the $N$-body results. At the smaller time-step, solution $II$ has slightly
higher values of $M_1$ than the simulation, and it also has an isolated peak
near $(M_1,M_2)=(M_0/2,0)$, with no parallel trace in the simulation. At
$\dW=1.7$, solution $II$ shows bigger deviations in the masses of $M_1$ and
$M_2$. Based on these findings, we adopt solution $I$ as our default option
for $P_{1,2}$. However, one should bear in mind that each of the three solutions
discussed above is an example of a solution that is fully consistent with the
EPS conditional mass functions.


\subsection{A More Realistic Model?}
\label{sec:the-solut}

The solution in terms of delta-functions, \equ{define_f}, is motivated by the
work of LC93 and by results from $N$-body simulations. We find that the
$P_{1,2}$ extracted from the Millennium simulation indeed approaches a narrow
function when $\dW\rightarrow0$. Nonetheless, it should be noted that the
delta-function solution is not the only possible solution for EPS even when
$\dW\rightarrow0$. We do find other EPS solutions with a broad $P_{2|1}$.
The delta-function treatment is simple, though it has its limitations. For
the finite time-steps used, the actual width of the distribution in the
$N$-body simulation is finite, not zero. With the optimal time-step for
reconstructing merger trees, $\dW\sim0.1$ (ND08), the delta-function solution
is accurate within EPS, but it is not such a good approximation to the
$N$-body merger trees.

A different approach might be to seek a solution that can be used with any
time-step $\dW$ in a self-consistent way, namely it should keep the same when
using $k$ time-steps of $\dW_1$ or one time-step of $\dW=k\times\dW_1$. Motivated
by ND08, we try
\begin{equation}
P_{2|1}(M_2|M_1,M_0,\dW) = P_1(M_2|M_a,\dW) \;,
\end{equation}
where $M_a=f_2(M_1)$ as defined in \equ{define_f}. This solution is fully
consistent with $\ptot$ for small enough $\dW$ as it approaches a delta function.
However, for big time-steps it shows some deviations from the theoretical
$\ptot$, depending on the specific solution adopted for $P_{1,2}$. This
solution is not practical for our applications because it does not fit accurately
the shape of $P_{1,2}$ as obtained from many small time-steps of solution $I$.
We mention it here because it is close to solution $II$ even for big time-steps.

For completeness, the explicit expression for the merger rate in this case is
\begin{eqnarray}
\lefteqn{ P_{1,2}(M_1,M_2|M_0,\dW) = \frac{1}{2\pi} \frac{M_0M_a}{M_1M_2}
\frac{(\dW)^2}{(\dS_1 \dS_2)^{1.5}} } \\ \nonumber & &
\!\!\!\!\!\!\!\!\!\!\!\!\!
\times\!\exp\!\left[\!-\frac{(\dW)^2}{2}\!\left(\frac{1}{\dS_1}\!+\!\frac{1}
{\dS_2} \right) \right] \left| \frac{\dd S(M_1)}{\dd M_1}\right|
\left| \frac{\dd S(M_2)}{\dd M_2}\right| .
\end{eqnarray}
where $\dS_1\!=\!S(M_1)\!-\!S(M_0)$ and $\dS_2\!=\!S(M_2)\!-\!S(M_a)$.


\section{Practical Applications}
\label{sec:implic}

We next implement the method outlined above for EPS merger rates,
specifically solution $I$, to compute several quantities concerning the
clustering of dark-matter haloes, which are of practical interest in
the studies of galaxy formation.


\subsection{Major and Minor Merger Rates for a Given Halo ${\bf M_0}$}

\begin{figure}
\centerline{ \hbox{ \epsfig{file=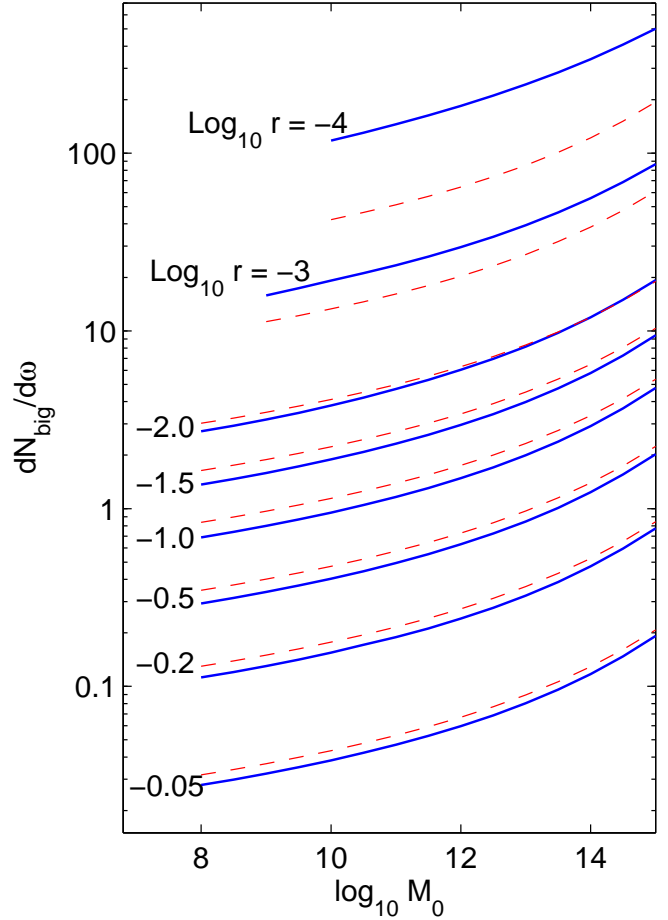,width=9cm} }}
\caption{The number of merger events with mass ratio $>r$,
per unit ``time" $\dd\omega$, for a given final halo mass $M_0$
(in units of $\hmsun$).
The value of $r$ associated with each pair of curves is indicated.
The solid curves describe the results of our EPS model.
The dashed curves are the results of the LC93 formula,
assuming that the main progenitor is more massive than $0.5M_0$
and $f_2^{LC}=M_0-M_1$.
}
  \label{fig:N_major}
\end{figure}

We first compute the probability that a halo of mass $M_0$ has undergone
within the last time-step $\dW$ a merger event that includes the main-progenitor
$M_1$ and another progenitor of mass $M_i>rM_1$ ($i\geq2$). Using our
definition for merger rate, section \ref{sec:merger-rates}, this can be written
as
\begin{eqnarray}
\lefteqn{ \frac{\dd N_{\rm big}(r,M_0)\ }{\dd\omega} = } \\ \nonumber & &
\frac{\dd}{\dd\omega} \sum_{i=2}^{\infty}
\int_{M_i>M_1r} P_{1,i}(M_1,M_i|M_0,\dW) \dd M_1 \dd M_i \;.
\end{eqnarray}
We use \equ{P_1i} in order to integrate over $M_i$ and obtain
\begin{eqnarray}
\frac{\dd N_{\rm big}(r,M_0)\ }{\dd\omega} =
\sum_i \int_{f_i(M_1)>M_1r}\frac{\dd P_{1}(M_1|M_0)}{\dd\omega}  \dd M_1 \;.
\end{eqnarray}
\Equ{dptot_dw} provides the derivative of $P_1$ with respect to $\omega$
for any $r>0$. This rate is independent of redshift as it is expressed in
terms of the self-invariant time variable $\omega$. When needed in units of
time, one should multiply the above expression by $\dot{\omega}$. A useful
approximation for $\dot{\omega}$ (from ND08) is
\begin{equation}
\dot{\omega} = -0.0470 \left[ 1+z+0.1(1+z)^{-1.25}\right]^{2.5}\; h_{73}\; {\rm Gyr}^{-1} \;,
\end{equation}
where $h_{73}$ is the Hubble constant in units of $73\kms$. This approximation
is valid for the $\Lambda$CDM cosmology used here, with
$(\Omega_m,\Omega_\Lambda)=(0.25,0.75)$, to better than $0.5\%$ at all
redshifts.

\Fig{N_major} shows results for $\dd N_{\rm big}(r,M_0) / \dd\omega$. For
example, we read that $\dd N_{\rm big}(0.3,10^{12}\hmsun) / \dd\omega \sim
0.65$, which means that a halo of mass $10^{12}\;\hmsun$ has undergone on
average $0.65$ major mergers of $r>0.3$ per unit of $\omega$. Multiplying
by $\dot{\omega}$ at $z=0$ gives $0.04$ major mergers per Gyr. At $z=3$ it
yields $\sim 1$ such mergers.
The number of minor mergers, with $10^{-4}<r<0.3$, is drastically higher;
a $10^{12}\hmsun$ halo has $\sim10$ such minor mergers per Gyr at $z=0$,
and $\sim 250$ such events per Gyr at $z=3$.

Figure \ref{fig:N_major} also shows the number of merger events as derived
from the formula of LC93, and assuming that the main progenitor is more
massive than $M_0/2$. The LC93 approach is interpreted here as $f_2^{\rm
LC}(M_1)=M_0-M_1$. The error due to their assumption is $\sim20\%$ for major
mergers, and it becomes as large as a factor of $\sim3$ at $r\sim10^{-4}$. We
emphasize that this is true for our default solution $I$. It is possible that
another EPS solution may be somewhat
closer to the LC93 results, but the discrepancy of
the LC93 estimates for minor mergers is likely to remain large.


\subsection{Growth Rate of a Halo ${\bf M_0}$ due to Major Mergers}

\begin{figure}
\centerline{ \hbox{ \epsfig{file=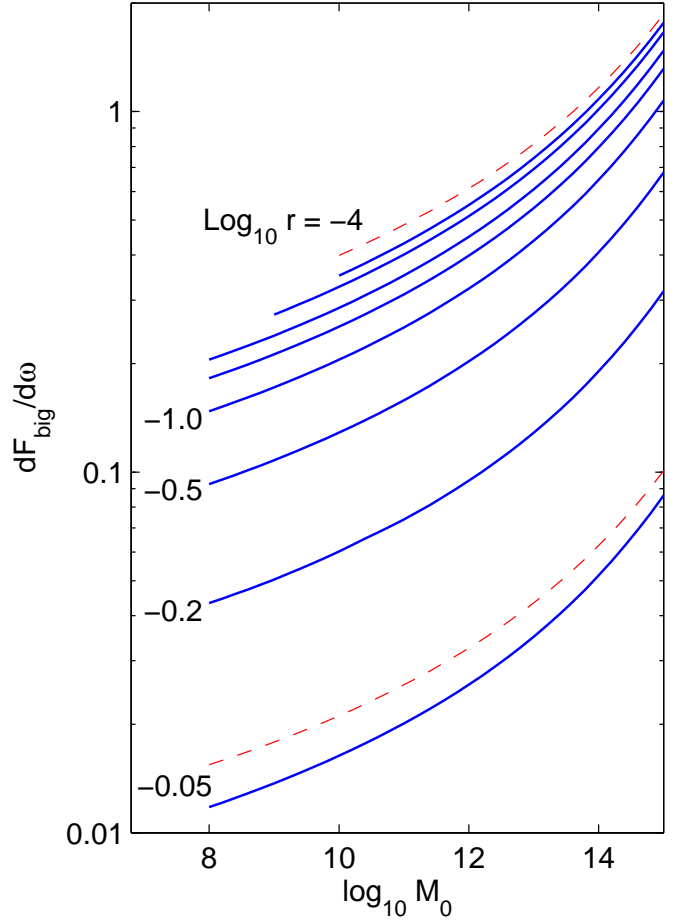,width=9cm} }}
\caption{The mass fraction added to a halo of mass $M_0$
by mergers with progenitors of mass ratio $>r$,
The values of $r$ are the same as in \fig{N_major}.
The solid curves describe our EPS model.
The dashed curves refer to LC93; they are plotted only for the extreme
values of $r$.
}
  \label{fig:F_major}
\end{figure}

As a second example we compute the average mass fraction added to a halo
by merger events with mass ratio greater than $r$,
\begin{eqnarray}
\lefteqn{ \frac{\dd F_{\rm big}(r,M_0)\ }{\dd\omega} = } \\ \nonumber & &
\frac{\dd}{\dd\omega} \sum_{i=2}^{\infty} \int_{M_i>M_1r}
P_{1,i}(M_1,M_i|M_0,\dW) \frac{M_i}{M_0} \dd M_1 \dd M_i \;.
\end{eqnarray}
As before, we can simplify the expression to
\begin{eqnarray}
\lefteqn{ \frac{\dd F_{\rm big}(r,M_0)\ }{\dd\omega} =  }
\\ \nonumber & & \sum_i \int_{f_i(M_1)>M_1r}
\frac{\dd P_{1}(M_1|M_0)}{\dd\omega} \frac{f_i(M_1)}{M_0} \dd M_1 \;.
\end{eqnarray}
Results for $\dd F_{\rm big}(r,M_0)/\dd\omega$ are shown in \fig{F_major}.
As an example, $\dd F_{\rm big}(0.3,10^{12}\hmsun)/\dd\omega$ reads $\sim 0.2$.
This means that a halo of mass $10^{12}\;\hmsun$ has gained on average
$\sim 20\%$ of its mass by major mergers per unit of $\omega$. Multiplying
by $\dot{\omega}$ we get a growth rate of $1\%$ per Gyr by major mergers at
$z=0$, and $\sim 30\%$ at $z=3$. For this quantity the LC93 assumption leads
to similar errors of $\lsim 20\%$ for all mass ratios $r$.

\subsection{Merger Rates for a given $M_1$}

The number of haloes of mass $M_s$ that will merge with a halo of a given
mass $M_1$ ($M_s<M_1$) within the time-step $\dW$, with no restriction on
the descendant mass $M_0$, is
\begin{eqnarray}
\lefteqn{ \frac{\dd Q(M_s|M_1,z)}{\dd\omega} = } \\ \nonumber
& & \int_{M_1+M_s}^{M_1/x_1} \frac{\dd P_{1,s}(M_1,M_s|M_0,\dW)}{\dd \omega} \;
 \frac{\phi(M_0,z)}{\phi(M_1,z)} \; \dd M_0 \;,
\end{eqnarray}
where $\phi(M,z)$ is the Press-Schechter average comoving number density of
haloes of mass $M$ at redshift $z$, \equ{PS}. Using $P_{1,i}$ from \equ{P_1i}
we get
\begin{eqnarray}
\label{eq:Q}
\lefteqn{ \frac{\dd Q(M_s|M_1,z)}{\dd\omega} = } \\ \nonumber & & \sum_i \frac{\dd P_1(M_1|M_{0,i})}{\dd\omega} \frac{\phi(M_{0,i},z)}{\phi(M_1,z)} \left| \frac{\dd f_i}{\dd M_0} \right|^{-1} \; ,
\end{eqnarray}
where $M_{0,i}$ are the values of $M_0$ for which $M_s\!=\!f_i(M_1|M_{0,i})$.
Note that the derivative $\dd f_i/\dd M_0$ is with respect to $M_0$ rather
than $M_1$.

\begin{figure}
\centerline{ \hbox{ \epsfig{file=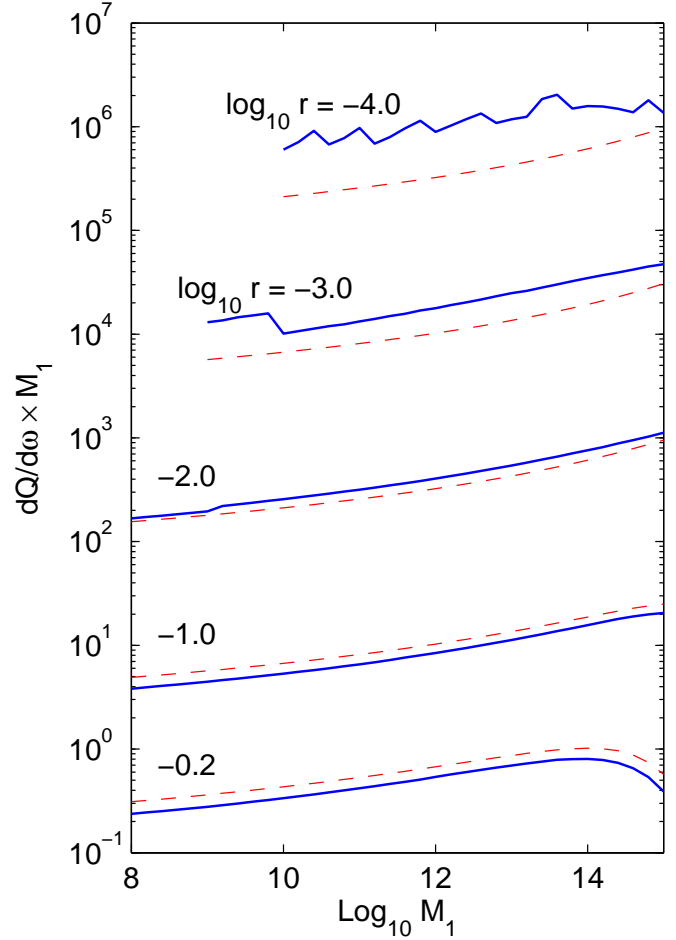,width=9cm} }}
\caption{The number of mergers of mass $M_s=rM_1$ with a given halo
of mass $M_1$, in an infinitesimal time-step $\dd\omega$.
The results are plotted for $z=0$.
The solid curves are results of the EPS solution $I$ given in
\se{specific_solut}.
The dashed curves follow the formula of LC93.
}
  \label{fig:merger_kernel}
\end{figure}

\Fig{merger_kernel} shows results for $\dd Q(M_s|M_1,z)/\dd\omega$. Unlike
the other quantities discussed above, $Q$ does depend explicitly on redshift
$z$, through the dependence of the sum in \equ{Q} on $z$. Nevertheless, for
major mergers (high $r$) this sum consists of only one term, so the $z$
dependence can be scaled out. Note also that $Q$ is not a smooth function,
due to the fact that $f_i(M_1)$ are always defined for an $M_1$ value that is
smaller than some threshold, in order to conserve mass (see the discussion
after \equ{IC_large}). \Fig{merger_kernel} displays in comparison the results
of the LC93 formula, showing deviations of $\sim30\%$ for major mergers, which
become as large as a factor of $\sim 3$ for a small mass ratio of
$r\sim 10^{-4}$.

\section{Creation and Destruction Rates of Haloes}
\label{sec:halo_creation}

In this section, we address the merger processes through
the creation and destruction rates of haloes.
We show that the results obtained here are consistent with the
Press-Schechter mass function, as they should be.
This can also serve as a sanity check for validating the
values of $Q$ computed above.

\citet{Benson05}
have attempted to use the Smoluchowski coagulation
equation for computing halo merger
rates. This equation evaluates the change in the number density of haloes due
to the competing processes of halo creation and destruction as a result of
mergers. Their formulae assume that each halo is formed by a binary merger
event, so halo formation is modeled as a 2-progenitor process. According to
the multiple progenitor theorem proved above, this approach cannot yield
correct results. A formulation of the Smoluchowski coagulation equation should
be therefore replaced with a different equation that takes
into account multiple mergers and accretion mass.
Consequently, the discrepancy found by \citet{Benson05} in the merger rate
formula of LC93 is not a discrepancy in the EPS formalism -- it simply reflects
the inaccuracy introduced in the LC93 formula by the assumption of binary
mergers. Despite this built-in error, the numerical merger rates by
\citet{Benson05} seem to be consistent with the Press-Schechter mass function.
This could be the result of using a mass grid cell size of $M_0/179$, which
scales with mass, while we found that multiple mergers become relevant only
for $M<10^{-3}M_0$.

The EPS formalism is, by construction, fully consistent with the
Press-Schechter mass function.  This can be expressed by
\begin{equation}
\phi(M,\omega+\dW) = \int_M^{\infty} \phi(M_0,\omega) \ptot(M|M_0,\dW) \dd M_0
\;.
\end{equation}
This equation indicates that any merger rate that is consistent with $\ptot$
must be consistent with the way $\phi$ varies in time. This implies that the
merger rates that were evaluated here should predict the correct rate of
change of $\phi$ when implemented using halo creation and destruction terms.

Taking into account the multiple progenitors and the accreted mass,
the time derivative of $\phi(M,z)$ is connected to $Q$ via the
equation
\begin{eqnarray}
\label{eq:coag}
\nonumber
\lefteqn{ -\frac{\dd \phi(M,z)}{\dd\omega} = \lim_{\Delta
M\rightarrow0, \;\; \dW\rightarrow0} \frac{1}{\dW\;2\Delta M} \bigg[ } \\
\nonumber
\lefteqn{ \int_{M-\Delta M}^{M+\Delta M} \phi(M_0,z) \dd M_0
\int_0^{M-\Delta M} P_1(M_1|M_0,\dW) \ \dd M_1 } \\
\nonumber
\lefteqn{ -\!\int_{M+\Delta M}^{\infty} \phi(M_0,z) \dd M_0 \int_{M-\Delta
M}^{M+\Delta M} P_1(M_1|M_0,\dW) \ \dd M_1 \bigg] } \\
\lefteqn{ -\!\int_M^\infty \frac{\dd Q(M|M_1,z)}{\dd \omega} \phi(M_1,z)
\dd M_1 \;. }
\end{eqnarray}
The first term corresponds to the creation of new haloes inside the mass bin
$[M-\Delta M,\; M+\Delta M]$ as arising from the
main-progenitor growth rate. The second term computes the number of haloes that
leave this bin for the same reason. We note that each of
these terms diverges for small $\Delta M$, but their sum remains constant.
The third term that involves $Q$ is the number of haloes
that leave the mass bin by merging with bigger haloes.
We have verified that this formula yields self-consistent results
by computing it term by term.
However, due to the numerical limitations of computing $Q$ in
only discrete points, we get an accuracy that is on the order of few percents
in the integral of $Q$.


\section{A Monte-Carlo Algorithm for EPS Merger Trees}
\label{sec:algo}

\begin{figure*}
\centerline{\psfig{file=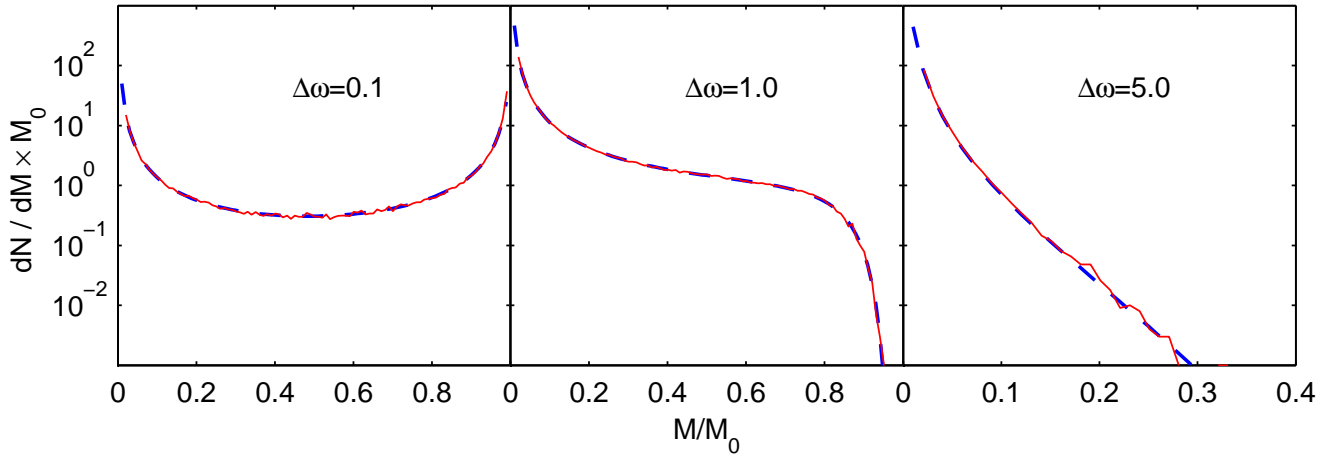,width=280mm,bbllx=1mm,bblly=118mm,bburx=210
mm,bbury=175mm,clip=}}
\caption{The number of progenitors,
\equ{dNdM}, using $M_0=10^{13}\; \hmsun$ and three different values of $\dW$.
The results based on $10^5$ random realizations with intrinsic
time-step $\dW=0.005$ are shown as solid (red) thin lines.
The dashed line is the theoretical prediction \equ{dNdM}.
}
\label{fig:mc_trees}
\end{figure*}

Using the specific analytical solution obtained in \se{specific_solut},
one can construct full merger trees. The algorithm is conceptually simple,
and can be summarized as follows:
\begin{itemize}
  \item Define a reference halo with mass $M_0$ at $\omega_0$.
  \item Choose a time-step $\dW$ (not necessarily small).
  \item Draw a random main-progenitor mass $M_1$,
using the distribution $P_1(M_1|M_0,\dW)$.
  \item Compute the value of $M_i$ ($i\geq2$) using $M_i=f_i(M_1|M_0,\dW)$,
for every value of $i$,
until the desired mass resolution is achieved.
  \item Repeat the above procedure for each progenitor $M_i$,
where $M_0$ is replaced by $M_i$.
\end{itemize}
This general algorithm can be used with any variant of the solutions
presented in \se{specific_solut}. An advantage of this algorithm is
that all the tree quantities can be computed analytically. Another advantage
over other algorithms is that its accuracy within EPS is in principle unlimited
--- it solely depends on the accuracy of the $f_i$ used. The algorithm can be
applied with time-steps that are not small, but the procedure is simpler when
using small time-steps so $\ptot$ is linear in $\dW$.

\Fig{mc_trees} shows results from EPS merger-tree realizations using our
algorithm based on solution $I$. These results demonstrate the high accuracy
of the generated trees.


\section{Summary and Discussion}
\label{sec:discuss}

We presented a rigorous method for computing dark-matter merger rates and
merger trees that obey the halo progenitor mass function of the EPS formalism
at any redshift. This corrects apparent inconsistencies within EPS
\citep{Lacey93, Benson05}. Our method conserves mass, in the sense that the
sum of the progenitor masses does not exceed the mass of the product halo.
This method translates the problem of constructing merger trees to solving
a differential equation. Different choices of initial conditions correspond
to different types of merger trees. This method enabled us to span the set
of solutions for merger rates within EPS, and to pick up a specific solution
whose merger trees are a good fit to $N$-body results. The same method can be
implemented with any conditional mass function beyond EPS, e.g., as extracted
from $N$-body simulations or from an ellipsoidal-collapse model.

Our main result is an accurate derivation for the merger rate of
dark-matter haloes, which differs from the classical result of \citet{Lacey93}.
This is due to our finding that within the EPS formalism, a merger event typically
involves many progenitors in a time-step, even when this time-step is infinitely
small, as opposed to the binary mergers assumed in previous works.
Our corrected results
differ from those derived by \citet{Lacey93} especially in the number of minor
merger events, while other quantities deviate only at the level of $20\%$. We
compute a few useful variants of the merger-rate formula, such as the number
of mergers for a given descendant halo, the mass fraction added by mergers, and
the merger rate per progenitor halo. These examples span many applications for
galaxy formation models. We also verified that the merger rates derived here are
fully consistent with the evolution of the Press-Schechter mass function, in
terms of counting the creation and destruction of haloes within the coagulation
equation.

We have shown that the merger rates derived here fit the results
of $N$-body simulations better than the early results of \citet{Lacey93}.
However, as discussed in \citet{Neistein08}, the merger rates
from $N$-body simulations may suffer from intrinsic inconsistencies at
the level of a few tens of percents due to non-Markov effects.
Keeping this in mind, it is tempting to compare our EPS merger rates
with other studies of merger-rates extracted from $N$-body simulations \citep[e.g.][]{Fakhouri07,Stewart07}.
For example, it is likely that our EPS results are in better agreement
with the $N$-body results than the EPS results presented by
\citet{Fakhouri07}; their EPS merger rates are underestimates at low mass ratio
of $r\sim 10^{-3}$. This better agreement is similar to what we find
here based on the merger-rates of \citet{Neistein08}.

The concept of multiple mergers in the limit of small time-steps, proven here to
be valid in EPS, deserves further attention. Recent studies indicate that
this might be true in $N$-body simulations when the time-steps used
are finite
\citep[e.g.][]{Fakhouri07,Neistein08}.
However,
in an $N$-body system, every merger event can be broken into a sequence
of binary mergers once the time step is short enough.
This implies that the $N$-body system is not a pure Markov process ---
the binary mergers are a non-Markov feature.
This non-Markov feature reflects a correlation between the successive
mergers. A correlation of this sort may be introduced, for example, by the
progenitors being part of a cosmic-web filament feeding a bigger halo,
where they merge in as a coherent group.
When we impose a Markov model to describe the $N$-body mergers, i.e. a
model that ignores any correlations,
this correlated sequence of  binary mergers is forced to
appear as a multiple merger.
It would be interesting to verify this interpretation of the relation
between the EPS and $N$-body mergers
by testing whether most of the $N$-body merger events are indeed part
of a correlated sequence.

If multiple-merger events are a common phenomena in EPS,
then the merger rates are not
defined in a unique way, as the counting method by which progenitors
are ordered to merge with each other may affect the merger-rate results.
Here we have chosen a simplified approach, where all the progenitors are
assumed to merge with the most massive progenitor and not with one another.
Clearly, other methods of counting may be applied.
This issue may be examined in detail using an $N$-body simulation,
where the multiple mergers can be broken into a sequence of binary events
once the time steps are made small enough.
It should be noted that
the conditional mass function of progenitors as extracted from $N$-body
simulations cannot be reconstructed by a Markov process \citep[see][]{Neistein08}.
This means that there is no accurate expression for this mass function at small
time-steps that can reproduce the mass function at high redshift,
namely separated from the present by a large time-step
\citep[but see][for an approximation]{Cole08}.
Further effort is needed in order to understand this issue in
$N$-body simulations.

A conditional mass function that is based on the ellipsoidal collapse
model has been used recently for generating merger trees \citep{Moreno07}
and for computing merger rates \citep{Zhang08}.
The use of the ellipsoidal model is partly motivated by its earlier success,
over the spherical model used by Press-Schechter,
in reproducing the (un-conditional) mass function of haloes in
$N$-body simulations \citep{Sheth02}. The method developed in this paper
can be easily generalized to utilize the ellipsoidal collapse model.
The results should be compared to our EPS predictions and to the $N$-body
results.  As a first step, it should be interesting to evaluate the
level of accuracy in previous studies due to the binary-merger assumption
by computing the average number of progenitors per merger event.

The algorithm we provide for generating merger trees has several
advantages as follows:
\begin{itemize}
  \item It is fully consistent with the EPS conditional mass function of
progenitors.
  \item The relevant statistics can be described analytically,
including those concerning the main-progenitor history and the merger rates.
  \item This algorithm was chosen, out of the many options that are consistent
with EPS, to provide best fit to $N$-body simulations.
  \item The constructed merger trees conserve mass, in the sense that
the total mass in progenitors does not exceed the descendant halo mass.
\end{itemize}
These are significant improvements over previous algorithms that follow EPS
\citep{Cole91, Kauffmann93, Sheth99, Somerville99a, Cole00, Hiotelis06},
which makes the new algorithm a useful tool for analytic and
semi-analytic modeling of galaxy formation.
Still, the non-EPS algorithms that are empirically tuned to match
$N$-body simulations \citep{Parkinson08,Neistein08} may have advantages in
certain cases where the accuracy is important.


\section*{Acknowledgments}

We thank Andrew Benson, Shaun Cole, Vincent Desjacques,
Jorge Moreno, Ravi Sheth and Simon White for comments and
stimulating discussions.
This research has been supported by GIF I-895-207.7/2005,
by a French-Israel Teamwork in Sciences,
by the Einstein Center at HU, and by NASA ATP
NAG5-8218 at UCSC.

\bibliographystyle{mn2e}
\bibliography{eyal_2}

\begin{thebibliography}{}

\bibitem[\protect\citeauthoryear{{Benson}, {Kamionkowski} \&
  {Hassani}}{{Benson} et~al.}{2005}]{Benson05}
{Benson} A.~J.,  {Kamionkowski} M.,    {Hassani} S.~H.,  2005, \mnras, 357, 847

\bibitem[\protect\citeauthoryear{{Bond}, {Cole}, {Efstathiou} \&
  {Kaiser}}{{Bond} et~al.}{1991}]{Bond91}
{Bond} J.~R.,  {Cole} S.,  {Efstathiou} G.,    {Kaiser} N.,  1991, \apj, 379,
  440

\bibitem[\protect\citeauthoryear{{Cole}}{{Cole}}{1991}]{Cole91}
{Cole} S.,  1991, \apj, 367, 45

\bibitem[\protect\citeauthoryear{{Cole}, {Helly}, {Frenk} \&
  {Parkinson}}{{Cole} et~al.}{2008}]{Cole08}
{Cole} S.,  {Helly} J.,  {Frenk} C.~S.,    {Parkinson} H.,  2008, \mnras, 383,
  546

\bibitem[\protect\citeauthoryear{{Cole}, {Lacey}, {Baugh} \& {Frenk}}{{Cole}
  et~al.}{2000}]{Cole00}
{Cole} S.,  {Lacey} C.~G.,  {Baugh} C.~M.,    {Frenk} C.~S.,  2000, \mnras,
  319, 168

\bibitem[\protect\citeauthoryear{{Cuesta}, {Prada}, {Klypin} \&
  {Moles}}{{Cuesta} et~al.}{2007}]{Cuesta07}
{Cuesta} A.~J.,  {Prada} F.,  {Klypin} A.,    {Moles} M.,  2007,
  astro-ph/0710.5520

\bibitem[\protect\citeauthoryear{{Epstein}}{{Epstein}}{1983}]{Epstein83}
{Epstein} R.~I.,  1983, \mnras, 205, 207

\bibitem[\protect\citeauthoryear{{Fakhouri} \& {Ma}}{{Fakhouri} \&
  {Ma}}{2007}]{Fakhouri07}
{Fakhouri} O.,  {Ma} C.-P.,  2007, astro-ph/0710.4567

\bibitem[\protect\citeauthoryear{{Hiotelis} \& {Popolo}}{{Hiotelis} \&
  {Popolo}}{2006}]{Hiotelis06}
{Hiotelis} N.,  {Popolo} A.~D.,  2006, \apss, 301, 167

\bibitem[\protect\citeauthoryear{{Kauffmann} \& {White}}{{Kauffmann} \&
  {White}}{1993}]{Kauffmann93}
{Kauffmann} G.,  {White} S.~D.~M.,  1993, \mnras, 261, 921

\bibitem[\protect\citeauthoryear{{Lacey} \& {Cole}}{{Lacey} \&
  {Cole}}{1993}]{Lacey93}
{Lacey} C.,  {Cole} S.,  1993, \mnras, 262, 627 (LC93)

\bibitem[\protect\citeauthoryear{{Lacey} \& {Cole}}{{Lacey} \&
  {Cole}}{1994}]{Lacey94}
{Lacey} C.,  {Cole} S.,  1994, \mnras, 271, 676

\bibitem[\protect\citeauthoryear{{Ludlow}, {Navarro}, {Springel}, {Jenkins},
  {Frenk} \& {Helmi}}{{Ludlow} et~al.}{2008}]{Ludlow08}
{Ludlow} A.~D.,  {Navarro} J.~F.,  {Springel} V.,  {Jenkins} A.,  {Frenk}
  C.~S.,    {Helmi} A.,  2008, astro-ph/0801.1127

\bibitem[\protect\citeauthoryear{Moreno \& Sheth}{Moreno \&
  Sheth}{2007}]{Moreno07}
Moreno J.,  Sheth R.~K.,  2007, astro-ph/0712.3800

\bibitem[\protect\citeauthoryear{{Neistein} \& {Dekel}}{{Neistein} \&
  {Dekel}}{2008}]{Neistein08}
{Neistein} E.,  {Dekel} A.,  2008, \mnras, 383, 615 (ND08)

\bibitem[\protect\citeauthoryear{{Neistein}, {van den Bosch} \&
  {Dekel}}{{Neistein} et~al.}{2006}]{Neistein06}
{Neistein} E.,  {van den Bosch} F.~C.,    {Dekel} A.,  2006, \mnras, 372, 933

\bibitem[\protect\citeauthoryear{{Parkinson}, {Cole} \& {Helly}}{{Parkinson}
  et~al.}{2008}]{Parkinson08}
{Parkinson} H.,  {Cole} S.,    {Helly} J.,  2008, \mnras, 383, 557

\bibitem[\protect\citeauthoryear{{Press} \& {Schechter}}{{Press} \&
  {Schechter}}{1974}]{Press74}
{Press} W.~H.,  {Schechter} P.,  1974, \apj, 187, 425

\bibitem[\protect\citeauthoryear{{Sheth} \& {Lemson}}{{Sheth} \&
  {Lemson}}{1999}]{Sheth99}
{Sheth} R.~K.,  {Lemson} G.,  1999, \mnras, 305, 946

\bibitem[\protect\citeauthoryear{{Sheth} \& {Pitman}}{{Sheth} \&
  {Pitman}}{1997}]{Sheth97}
{Sheth} R.~K.,  {Pitman} J.,  1997, \mnras, 289, 66

\bibitem[\protect\citeauthoryear{{Sheth} \& {Tormen}}{{Sheth} \&
  {Tormen}}{2002}]{Sheth02}
{Sheth} R.~K.,  {Tormen} G.,  2002, \mnras, 329, 61

\bibitem[\protect\citeauthoryear{{Somerville} \& {Kolatt}}{{Somerville} \&
  {Kolatt}}{1999}]{Somerville99a}
{Somerville} R.~S.,  {Kolatt} T.~S.,  1999, \mnras, 305, 1

\bibitem[\protect\citeauthoryear{{Springel}, {White}, {Jenkins}, {Frenk},
  {Yoshida}, {Gao}, {Navarro}, {Thacker}, {Croton}, {Helly}, {Peacock}, {Cole},
  {Thomas}, {Couchman}, {Evrard}, {Colberg} \& {Pearce}}{{Springel}
  et~al.}{2005}]{Springel05}
{Springel} V.,  {White} S.~D.~M.,  {Jenkins} A.,  {Frenk} C.~S.,  {Yoshida} N.,
   {Gao} L.,  {Navarro} J.,  {Thacker} R.,  {Croton} D.,  {Helly} J.,
  {Peacock} J.~A.,  {Cole} S.,  {Thomas} P.,  {Couchman} H.,  {Evrard} A.,
  {Colberg} J.,    {Pearce} F.,  2005, \nat, 435, 629

\bibitem[\protect\citeauthoryear{{Stewart}, {Bullock}, {Wechsler}, {Maller} \&
  {Zentner}}{{Stewart} et~al.}{2007}]{Stewart07}
{Stewart} K.~R.,  {Bullock} J.~S.,  {Wechsler} R.~H.,  {Maller} A.~H.,
  {Zentner} A.~R.,  2007, astro-ph/0711.5027

\bibitem[\protect\citeauthoryear{{Zhang}, {Ma} \& {Fakhouri}}{{Zhang}
  et~al.}{2008}]{Zhang08}
{Zhang} J.,  {Ma} C.-P.,    {Fakhouri} O.,  2008, astro-ph/0801.3459

\end{thebibliography}

\appendix
\section{More Results}

This Appendix is a supplement to \se{implic}, for the benefit of
practitioners who desire to read out numerical values for merger rates from the figures. Figures 6-8 of \se{implic} present merger-rate quantities as a
function of halo mass for different given values of mass ratio $r$.
Here we plot the same merger rate quantities as a function of $r$ for
different values of mass. This way of presenting the merger rates emphasizes their simple scaling with halo mass and highlights the trends at small values of $r$.

\begin{figure}
\centerline{ \hbox{ \epsfig{file=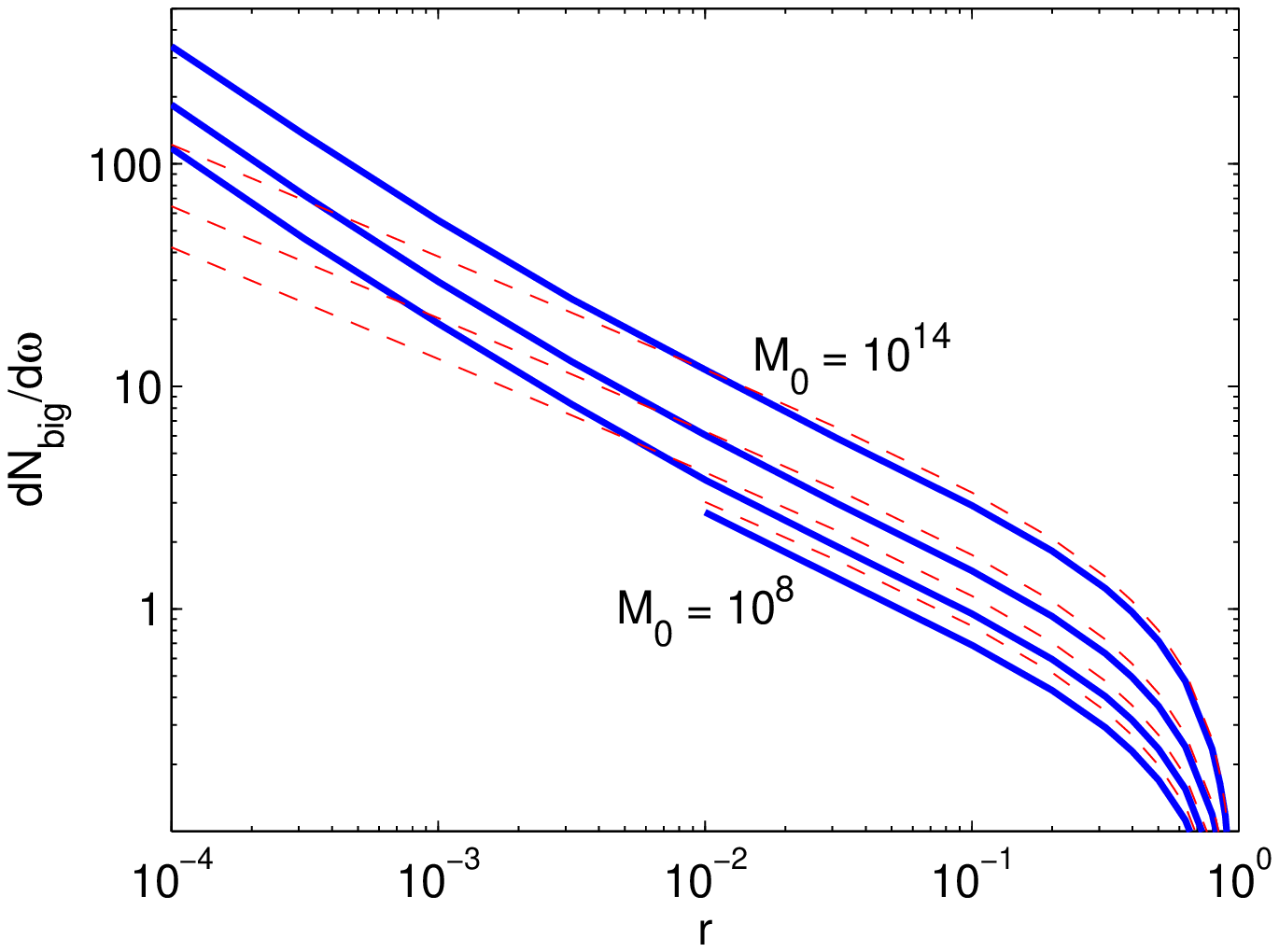,width=9cm} }}
\caption{The number of merger events with mass ratio $>r$,
per unit ``time" $\dd\omega$, for a given final halo mass $M_0$
(in units of $\hmsun$).
The values of $M_0$ are $10^8,\;10^{10},\;10^{12},\;10^{14}\; \hmsun$.
The solid curves describe the results of our EPS model while
the dashed curves are the results of the LC93 formula.
}
  \label{fig:N_major_app}
\end{figure}

\begin{figure}
\centerline{ \hbox{ \epsfig{file=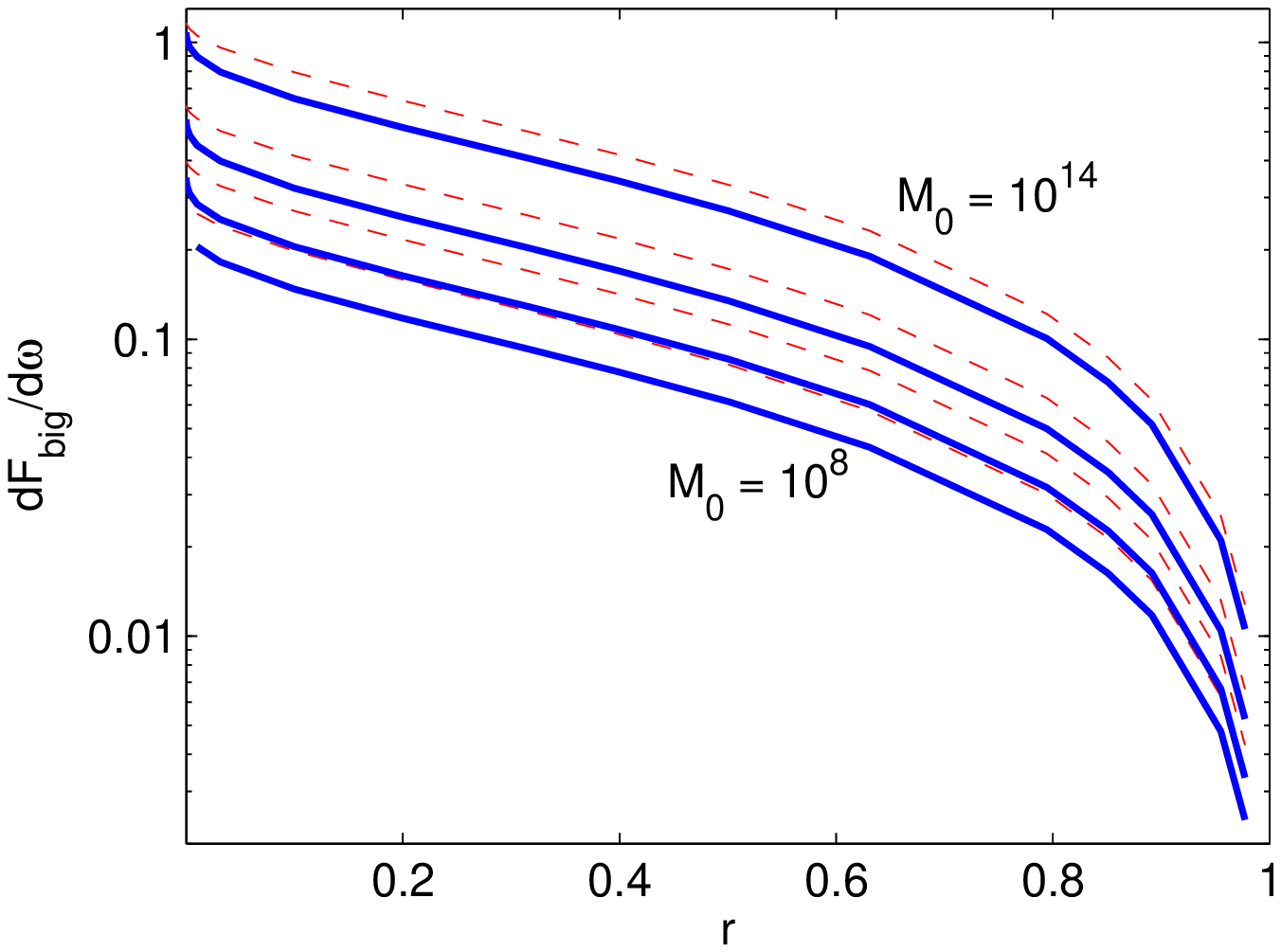,width=9cm} }}
\caption{The mass fraction added to a halo of mass $M_0$
by mergers with progenitors of mass ratio $>r$,
The values of $M_0$ are the same as in \fig{N_major_app}.
The solid curves describe our EPS model and
the dashed curves refer to LC93.
}
  \label{fig:F_major_app}
\end{figure}

\begin{figure}
\centerline{ \hbox{ \epsfig{file=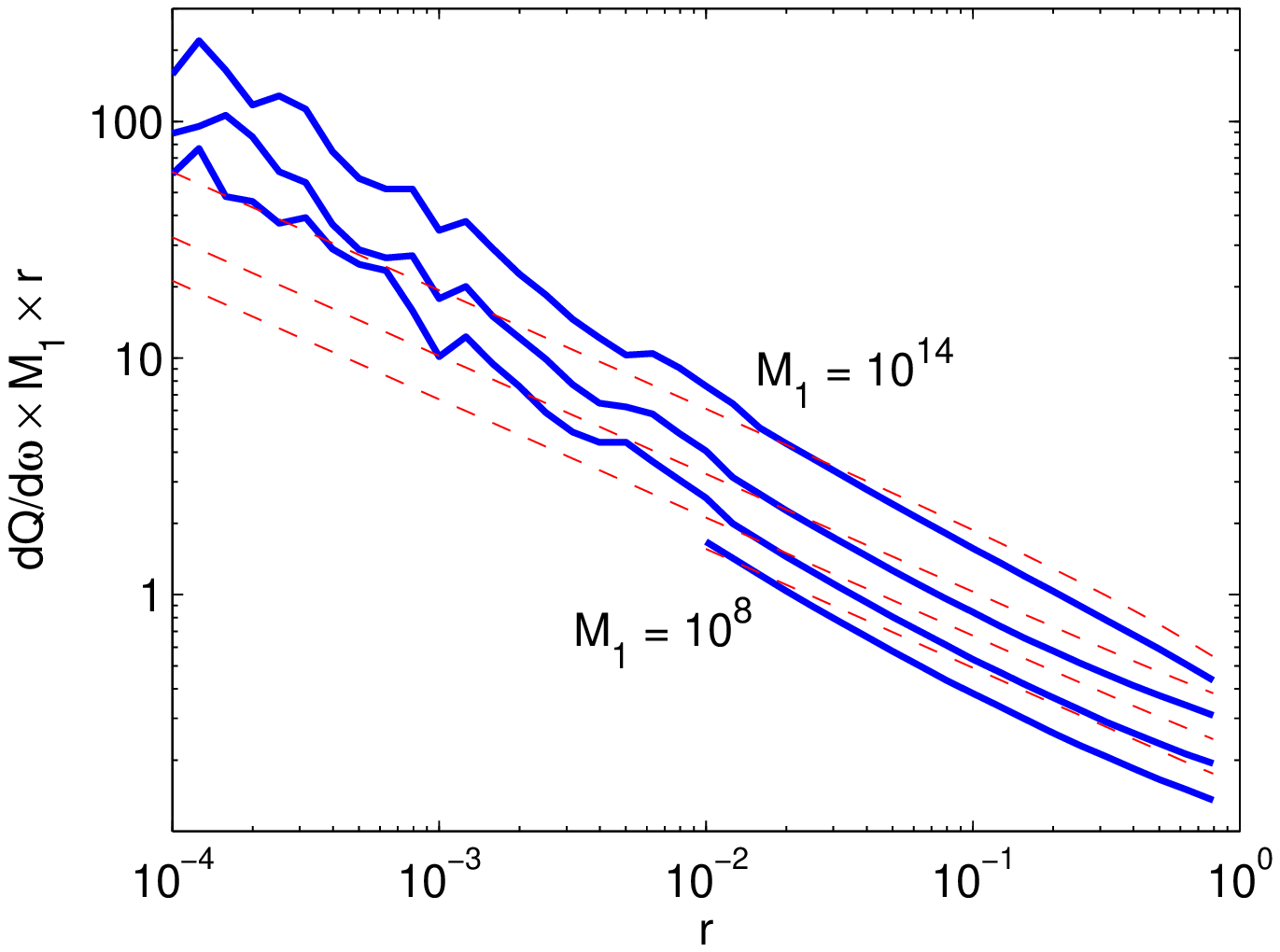,width=9cm} }}
\caption{The number of mergers of mass $M_s=rM_1$ with a given halo
of mass $M_1$, in an infinitesimal time-step $\dd\omega$.
The results are plotted for $z=0$, and for $M_1 = 10^8,\;10^{10},\;10^{12},\;10^{14}\; \hmsun$.
The solid curves follow the EPS solution $I$ given in
\se{specific_solut}, and
the dashed curves are obtained from the formula of LC93.
}
  \label{fig:merger_kernel_app}
\end{figure}


\label{lastpage}

\end{document}